\documentclass[prd,showpacs,aps,preprint]{revtex4}
\usepackage{graphicx}
\usepackage{dcolumn}
\usepackage{amsfonts}

\newcommand{\ep}{\epsilon}
\newcommand{\vep}{\varepsilon}

\newcommand{\td}{\tilde}

\newcommand{\beq}[1]{\begin{eqnarray}\label{#1}}
\newcommand{\eeq}{\end{eqnarray}}
\newcommand{\net}{{\cal N}=1}

\newcommand{\CN}[1]{{\cal N}=#1}

\begin{document}
\title{Integrable Spin Chain and Operator Mixing\\
in $\CN{1,2}$ Supersymmetric Theories} \vspace{0.2in}
\author{Xiao-Jun Wang$^1$}
\email{wangxj@ustc.edu.cn}
\author{Yong-Shi Wu$^{2,3}$}
\email{wu@physics.utah.edu} \affiliation{ \centerline{$^1$
Interdisciplinary Center for Theoretical Study}
\centerline{University of Science and Technology of China}
\centerline{AnHui, HeFei 230026, China} \centerline{$^2$
Department of Physics, University of Utah} \centerline{Salt Lake
City, Utah 84112, USA} \centerline{$^3$ Interdisciplinary Center
for Theoretical Study} \centerline{Chinese Academy of Sciences,
Beijing 100080, China}}

\begin{abstract}
We study operator mixing, due to planar one-loop corrections, for
composite operators in $D=4$ supersymmetric theories. We present
some ${\cal N}=1,2$ Yang-Mills and Wess-Zumino models, in which
the planar one-loop anomalous dimension matrix in the sector of
holomorphic scalars is identified with the Hamiltonian of an
integrable quantum spin chain with $SU(3)$ or $SU(2)$ symmetry,
even if the theory is away from the conformal points. This points
to a more universal origin of the integrable structure beyond
superconformal symmetry. We also emphasize the role of the
superpotential in the appearance of the integrable structure. The
computations of operator mixing in our examples by solving Bethe
Ansatz equations show some new features absent in $\CN{4}$ SYM.
\end{abstract}
\preprint{USTC-ICTS-03-13} \maketitle

\section{Introduction}

Whether the integrable structure that is abundant in two
dimensional field theories or statistical models could emerge in
four dimensions has been a fascinating topic on the research
frontier. Indeed, a (non-topological) Yang-Mills theory in four
dimensions should have very rich physics, so that it is hard to
believe it could be exactly solved as a whole the same way as in
two dimensions. On the other hand, one cannot either rule out the
possibility that an integrable structure appears in a subsector or
in a special limit. Actually, there have been quite a few
evidences for an integrable structure in the self-dual sector of a
Yang-Mills theory \cite{CGW82,D82,TW82,UN82,ADM86,B95,Popov} or in
the high energy limit of QCD \cite{L94,FK95,Kor03}.

Recently there have been revived interests in the integrable
structure in ${\cal N}=4$ superconformal Yang-Mills (SCYM) theory
in four dimensions, at least in the planar limit. One development
\cite{KL03} was to apply the integrable structure, originally
found in the QCD context \cite{Kor98}, to the computation of
anomalous dimensions of twist-two operators in ${\cal N}=4$ SCYM.
(This result was also obtained by means of other methods
\cite{DL02}.) Another strong evidence came from a recent one-loop
calculation of the anomalous dimensions of composite operators
\cite{MZ02}. (The interest in composite operators in the gauge
theory was mainly inspired by the proposal \cite{BMN} that at
least some of them could be viewed as the holographic dual of
string states in a special limit of the curved background
$AdS_5\times S^5$.) It turned out \cite{MZ02} that the planar
one-loop mixing matrix for the composite operators, that consist
of a string of scalar fields in the theory, is equivalent to the
Hamiltonian of an integrable spin chain with $SO(6)$ symmetry! One
may naturally ask whether a similar integrable structure could
appear in other Yang-Mills theories, or Wess-Zumino models, with
less supersymmetries or even with conformal symmetry broken. This
is an interesting issue, because one wants to know whether the
appearance of the integrable structure in the ${\cal N}=4$ SCYM is
related to the maximal superconformal symmetry that the theory
has.

In this paper, we present examples in which a similar integrable
structure survives deformation of the theory with less unbroken
supersymmetries. More concretely, we show that in some $\net$ (and
${\cal N}=2$) supersymmetric Yang-Mills theories in four
dimensions, the planar one-loop anomalous dimension matrix for
composite operators of holomorphic scalars is equivalent to the
Hamiltonian of an integrable quantum spin chain, even if the
theory is {\it away from} the conformal points. If we take a limit
in which the 't Hooft gauge couplings vanish, these models reduce
to Wess-Zumino models, and the integrable structure still survives
this limit. Our results indicate that the appearance of the
integrable structure in Yang-Mills theory in four dimensions
should have a more universal and profound origin, not restricted
only to theories with (maximal) superconformal symmetry. Moreover,
in our examples, the superpotential term is seen to play a crucial
role in determining the integrable structure.

Some of our models are deformation of orbifolding daughters of the
${\cal N}=4$ SCYM with the superpotential strength changed. In
their conformal phase, they are reduced to a quiver theory with a
product gauge group \cite{DM96} obtained by orbifolding the ${\cal
N}=4$ SCYM with a discrete subgroup of $SO(6)$ $R$-symmetry. We
therefore expect that there is a close relationship between
integrable spin chains in these $\net,2$ SYM and the $SO(6)$ chain
appearing in the ${\cal N}=4$ SCYM. Indeed it turns out that all
the integrable spin chains revealed by us in these $\net,2$ SYM is
closely related to a closed subsector of an $SO(6)$ chain; at
conformal points the parameters of the latter reduce to those of
the $SO(6)$ chain in ${\cal N}=4$ SCYM. However, the operator
mixing that results from diagonalizing the spin Halmitonian is
completely different, because of an interesting interplay between
the global symmetry index and the discrete index for gauge group
factors, which is absent in the $\CN{4}$ SYM.

The results for our $\CN{1,2}$ models, combined together, provide
a description of cascade breaking of the symmetry of the
integrable spin chains starting from the $\CN{4}$ SYM. This
motivates to make the conjecture that in {\it all} $\CN{1,2}$
orbifolded daughters \cite{KS98,LNV98} of the $\CN{4}$ SYM (with a
non-abelian global symmetry), with their 't Hooft and
superpotential couplings deformed away from the conformal points
while keeping the global symmetry of the superpotential, there is
always a non-trivial {\it integrable structure} in operator mixing
at planar one-loop level for (anti-) holomorphic composite
operators, at least for those consisting of purely scalars without
derivatives. Moreover, this integrable structure survives in the
resulting Wess-Zumino models when all 't Hooft couplings are sent
to zero.

The contents of this paper are as follows. In section 2 we
construct an $\CN{1}$ SYM model whose conformal phase is the
orbifolding limit of $\CN{4}$ SCYM. In section 3 we compute the
planar one-loop corrections to composite operators consisting of
scalar fields of the model. In section 4 we obtain the mixing
matrix for the renormalized composite operators, and show that the
anomalous dimension matrix in the (anti-)holomorphic sector is
equivalent to a Hamiltonian of an integrable $SU(3)$ spin chain.
We generalize our discussion to other $\CN{1}$ (and $\CN{2}$)
Yang-Mills and Wess-Zumino models in section 5, and devote section
6 to a brief summary. In two appendixes we present examples for
computing planar one-loop anomalous dimensions via solving the
Bethe ansatz equations of the $SU(3)$ quantum spin chain in
section 4 for our $\CN{1}$ model. This computation will explicitly
demonstrate how and why the operator mixing differs from the
$\CN{4}$ SYM case, though the quantum spin chain can be viewed as
a closed subsector of the latter.

\section{The ${\cal N}=1 $ model}

Various $\CN{1}$ superconformal gauge theories have been
constructed \cite{KS98,LNV98} via considering D3-branes at
orbifold singularities of the form $\mathbb{C}^3/\Gamma$ (with
$\Gamma$ a discrete subgroup of the $SO(6)$ R-symmetry of ${\cal
N}=4$ SYM) \cite{DM96,DGM97}. In AdS/CFT correspondence
\cite{M97}, the $\Gamma$ action is translated to an action
$AdS_5\times (S^5/\Gamma)$ with the $AdS$ part unaffected. So the
world-volume theory on $N$ D3-branes remains to be a conformal
field theory. With appropriately chosen $\Gamma$, supersymmetry is
broken down to $\CN{2},1$ or 0.

The simplest example for $\CN{1}$ SCYM is the case with
$\Gamma=\mathbb{Z}_3$ proposed in ref. \cite{KS98}. The gauge
group of the resulting theory has a product group
$U(N)^{(1)}\times U(N)^{(2)}\times U(N)^{(3)}$. We will use a
dsicrete index $A=1,2,3$ to label these $U(N)$ groups. The matter
in the theory consists of the bi-fundamental chiral superfields
\beq{1} 3\{(N,\bar{N},1)\oplus (\bar{N},1,N)\oplus
(1,N,\bar{N})\},
\eeq
and their anti-chiral partners (Fig. 1). Inside each pair of
circular brackets, we have the representations of the gauge
groups, while the overall factor $3$ in Eq. (\ref{1}) reflects an
$SU(3)$ global symmetry inherited from the SO(6) symmetry of the
${\cal N}=4$ SYM, that is broken down to $SU(3)$ by orbifolding.

\begin{figure}[hpbt]
\label{f1} \centering
\includegraphics[width=2.5in]{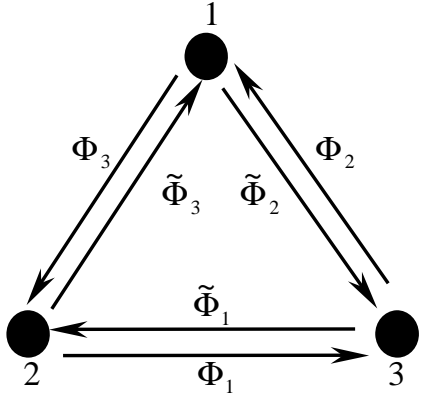}
\begin{minipage}{5in}
\caption{A quiver diagram for D3-branes on a
$\mathbb{C}^3/\mathbb{Z}_3$ orbifold. At the nodes we have
vectormultiplets in the gauge group indicated, while the arrows
connecting each pair of nodes correspond to the bi-fundamental
fields.}
\end{minipage}
\end{figure}

Therefore, each of the $U(N)$ groups is coupled to $N_f=3N$
fundamental chiral and $N_{\bar f}=3N$ anti-fundamental
anti-chiral matter. One may denote the bi-fundamental chiral
matter fields in Eq. (\ref{1}) as $\Phi^{a,(A\bar{B})}$. Here the
pair of indices, $(A,\bar{B})$, labels which two of the gauge
groups that the field is coupled to; and the index $a=1,2,3$ is
that for the representation ${\bf 3}$ of $SU(3)$. To simplify the
notation, we will write $\Phi^a_A=\Phi^{a,(B\bar{C})}$ with
$(A,B,C)$ being a cyclic permutation of (1,2,3); and denote their
Hermitian conjugate as $\td{\Phi}_{\bar A}^{\bar
a}=(\Phi_A^a)^\dag$ (Fig. 1), with the tilde symbol suppressed
when this is no confusion.

With the matter content~(\ref{1}), one can construct an $\net$
gauge theory, which is not necessarily conformal invariant by
allowing unequal gauge couplings and/or a more general
superpotential. In the standard $\CN{1}$ superfield formalism, the
Lagrangian of our model reads
\beq{2}{\cal L}&=&\frac{1}{4}\int
d^4\theta \sum_A {\rm tr}\{(\Phi_A^a)^{\dag}e^{V_B}\Phi_A^a
e^{-V_C}\} \nonumber \\ &+&\frac{1}{4}\int d^2\theta {\rm
tr}(W_AW_A+{\rm h.c.}) +\frac{1}{2}\int d^2\theta(\cal{W}+{\rm
h.c.}).
\eeq
In the first term, we assumed that $(A,B,C)$ is a cyclic
permutation of $(1,2,3)$. Here to deform the theory obtained by
orbifolding the $\CN{4}$ SYM, we allow the gauge couplings, $g_A$
with $A=1,2,3$, of the three $U(N)$ groups to be different from
each other,  and take the superpotential ${\cal W}$ to be a
generic $SU(3)$ invariant one:
\beq{3}
{\cal W}= \frac{h}{3}\ep_{abc} {\rm tr}\Phi_1^a\Phi_2^b\Phi_3^c.
\eeq
with $h$ {\it arbitrary}. So the global symmetry remains to be
$SU(3)$ after the deformation. If the three gauge couplings are
the same, our model also has a discrete $\mathbb{Z}_3$ symmetry
acting on the index $A$. The one-loop beta functions in this model
can be extracted from the NSVZ beta function \cite{NSVZ} as well
as the results in ref.  \cite{LS95}:
\beq{4}\beta_{\lambda_A}
&=&\frac{d\lambda_A}{d\ln\mu} = -\frac{\lambda_A^2}{4\pi^2}
(1-\frac{\lambda_A}{8\pi^2})^{-1}
\sum_{B\neq A}\gamma_B, \nonumber \\
\beta_h&=&\frac{d\lambda_h}{d\ln\mu} =\lambda_h\sum_A \gamma_A,
\eeq
where $\lambda_A=g_A^2N \; (A=1,2,3)$ are 't Hooft couplings for
the three $U(N)$ groups and $\lambda_h=|h|^2 N$; $\gamma_A$ are
anomalous dimensions of the three types of bi-fundamental scalar
fields. In large $N$ limit, the one-loop $\gamma_A$ are
\beq{5}\gamma_A=\frac{1}{2}\frac{d\ln{Z_{\phi_A}}}
{d\ln{\mu}}= \frac{1}{16\pi^2} (\sum_{B\neq
A}\lambda_B-2\lambda_h),
\eeq

In this paper we will work in the region with weak 't Hooft
couplings and small $\lambda_h$. Substituting Eq.~(\ref{5}) into
Eq. (\ref{4}) we have
\beq{6}\beta_{\lambda_A}&=&\frac{d\lambda_A}{d\ln\mu}
=- \frac{\lambda_A^2}{64\pi^4}(\sum_{B}\lambda_B
+\lambda_A-4\lambda_h)+O(\lambda^4),
\nonumber \\
\beta_h&=&\frac{d\lambda_h}{d\ln\mu}=\frac{\lambda_h}{8\pi^2}
(\sum_A \lambda_A - 3\lambda_h)+O(\lambda^3).
\eeq
There is a line of fixed points,
$\lambda_1=\lambda_2=\lambda_3=\lambda_h$, corresponding to the
orbifolded SCYM. Though the above equations cannot be exactly
solved we can consider the simplest case with
$\lambda_1=\lambda_2=\lambda_3$, denoted as $\lambda$, to
demonstrate some features for the flow. First from the flow
equations it is easy to see that if initially the three gauge
couplings coincide, then they will keep to do so during the
renormalization group flow. Secondly we can easily see that during
the flow one always has $\lambda_h=c\exp{(6\pi^2/\lambda)}$, with
an arbitrary positive integral constant $c$. So for fixed value of
$c$, both the 't Hooft and the superpotential couplings, $\lambda$
and $\lambda_h$, respectively, possess a fixed point
$\lambda=\lambda_h=\lambda_*(c)$, whose value is determined by the
root of the transcendental equation $x=c\exp{(6\pi^2/x)}$.  At the
fixed point, the theory is an interacting conformal field theory
at the quantum level. (The theory is no longer asymptotically free
for finite $\lambda_*(c)$). When $c$ varies one obtains a line of
RG fixed points. When away from the conformal points, at infra-red
the theory will be dominant by gauge interactions if
$\lambda>\lambda_*,\; \lambda_h<\lambda_*$ and by superpotential
interactions if $\lambda<\lambda_*,\; \lambda_h>\lambda_*$. In the
following we will calculate the operator mixing for small values
of the couplings $\lambda_A$ and $\lambda_h$.

\section{Matrix of Renormalization Factors }

We will study one-loop renormalization of composite operators,
which are a product of bi-fundamental scalars $\phi_A^a$, the
scalar component of the supermultiplets $\Phi_A^a$, and their
Hermitian conjugate $\phi_{\bar A}^{\bar a}=(\phi_A^a)^\dag$,
without derivatives:
\beq{7}
{\cal O}^{I_1...I_L}={\rm tr}\; \phi^{I_1}...\phi^{I_L},
\eeq
where each index $I_l\; (1\le l \le L)$ stands for a pair of
indices $(a_l,A_l)$ or $\bar{a}_l,\bar{A}_l$. The index $I_l$ for
a fixed $l$ can take 18 different values. The fields at the
right-hand side are all taken to be at the same space-time point.
These operators form (reducible) $SU(3)$ tensors with $L$ indices.
In particular, this class of composite operators includes purely
chiral (or holomorphic) operators which are a product of
$\phi_A^a$'s only (with no indices of type $\bar{a}$). In physics
one is restricted to composite operators that are {\it gauge
invariant} after taking the trace over the gauge-group matrix
indices. To compute operator mixing, it is better to work with the
composite operators before taking the trace. We note that gauge
invariance strongly constrain possible choice of the index
sequence $(I_1,\cdots,I_L)$. (See below.) In general, the scalar
operators~(\ref{7}) mix under renormalization, and the
renormalized operators with definite dimension are linear
combinations of the bare operators. If we specify a particular
operator basis, ${\cal O}^I$ with $I$ being a sequence
$I_1,\cdots,I_L$, the matrix $Z^I_J$ of renormalization factors in
this basis,
\beq{8}{\cal O}^I_{\rm ren}={Z^I}_J{\cal O}^J,
\hspace{0.5in} (I=1,...,18^L),
\eeq
is defined by the requirement of canceling UV divergences in the
correlation functions \beq{9}\langle
Z_{\phi_{I_1}}^{1/2}\phi_{I_1}(x_1)... Z_{\phi_{I_L}}^{1/2}
\phi_{I_L}(x_L) \; {\cal O}^{J_1...J_L}_{\rm ren}(x)\rangle,
\eeq
These $Z^I_J$ depend on the UV cutoff $\Lambda$ as well as various
couplings of the theory (in the large-$N$ limit), and form a $18^L
\times 18^L$ matrix. In rest of this section, we will compute the
$\delta Z^I_J \equiv Z^I_J - \delta^I_J$ due to planar one-loop
diagrams, using the component field formalism as in ref.
\cite{MZ02}. There are three types of planar one-loop diagrams, as
shown in Fig. 2, that contribute to the correlations~(\ref{9}). We
will choose the Fermi-Feynman gauge for gauge boson propagators,
in which the anomalous dimension of a single scalar field is
already given by Eq.~(\ref{5}).
\begin{figure}[hptb]
\label{f3} \centering
\includegraphics[width=4in]{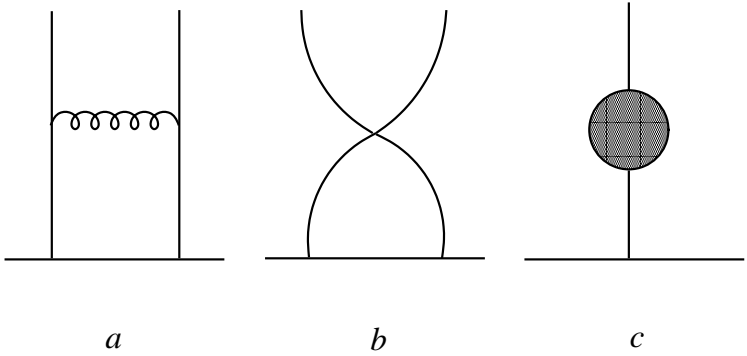}
\begin{minipage}{5in}
\caption{One-loop planar diagrams on correlation
functions~(\ref{9}).}
\end{minipage}
\end{figure}

The one-loop self-energy diagram, Fig. 2c, leads to the wave
function renormalization of scalar fields, which can be directly
read off from Eq.~(\ref{5}). Half of the self-energy correction in
correlation functions (\ref{9}) are cancelled by wave function
renormalization of the external legs. The counterterms that cancel
the remaining divergences are given by
\beq{10}
\delta Z^{(c)\cdots  J_lJ_{l+1}\cdots}_{\quad \cdots
I_lI_{l+1}\cdots } = -\frac{1}{2}(\gamma_{A_l}+\gamma_{A_{l+1}})
\ln{\Lambda}\; \delta^{J_l}_{I_l}\delta_{I_{l+1}}^{J_{l+1}},
\eeq
where each $I_l$ or $J_l$ stands for a pair of indices $(i_l,A_l)$
or $(j_l,B_l)$, respectively. Note that the contribution of this
diagram to the $Z$-matrix is diagonal in the indices $I_l, \;J_l$
and in $I_{l+1},\;J_{l+1}$.

As for the contributions from Fig. 2a and 2b, we have to
distinguish between two different cases: i)  two nearest-neighbor
scalar fields both are holomorphic (or anti-holomorphic), namely
${\cal O}\sim \;\cdots \phi^{a_l}_{A_l}
\phi^{a_{l+1}}_{A_{l+1}}\cdots $ (or $\cdots
\phi^{\bar{a}_l}_{\bar{A}_l}
\phi^{\bar{a}_{l+1}}_{\bar{A}_{l+1}}\cdots $); ii) one of them is
holomorphic and the other anti-holomorphic, i.e. ${\cal O}\sim
\cdots \phi^{a_l}_{A_l} \phi^{\bar{a}_{l+1}}_{\bar{A}_{l+1}}
\cdots $ (or $\cdots {\phi}^{\bar{a}_l}_{\bar{A}_l}
\phi^{a_{l+1}}_{A_{l+1}} \cdots $). If the nearest-neighbor pairs
in a composite operator ${\cal O}$ all belong to the above case
i), we call it a {\it holomorphic} (or {\it anti-holomorphic},
respectively) operator. Otherwise, if the above case ii) happens
to one nearest-neighbor pair, the composite operator is called
{\it non-holomorphic}.

The correction due to gauge boson exchange, Fig. 2a, contributes
to the $Z$-matrix associated with holomorphic operators:
\beq{11}
\delta Z^{(a)\cdots J_lJ_{l+1} \cdots}_{\quad \cdots
I_lI_{l+1}\cdots} =\frac{\lambda_{A_l^-}} {16\pi^2} \ln{\Lambda}\;
\delta_{I_l}^{J_l}\delta_{I_{l+1}}^{J_{l+1}},
\eeq
where $A_l^\pm=(A_l\pm 1)\;{\rm mod}\;3$. For anti-holomorphic
operators, one just replaces $A_l^-\to \bar{A}_l^+$. Meanwhile,
the same one-loop diagram, Fig. 2a, yields the following
$Z$-matrix for non-holomorphic operators:
\beq{12}
\delta Z^{(a)\cdots J_l \bar{J}_{l+1}\cdots}_{\quad \cdots I_l
\bar{I}_{l+1}\cdots} &=& \frac{\lambda_{A_l^-}}{16\pi^2}
\ln{\Lambda}\; \delta_{I_l}^{J_l}
\delta_{\bar{I}_{l+1}}^{\bar{J}_{l+1}},
\nonumber \\
\delta Z^{(a)\cdots {\bar{J}}_l J_{l+1} \cdots}_{\quad \cdots
{\bar I}_l I_{l+1}\cdots} &=& \frac{\lambda_{\bar{A}_l^+}}
{16\pi^2} \ln{\Lambda}\; \delta_{\bar{I}_l}^{\bar{J}_l}
\delta_{I_{l+1}}^{J_{l+1}}.
\eeq
Note that here an overall factor of $\delta_{A_l\bar{A}_{l+1}}$ or
$\delta_{\bar{A}_lA_{l+1}}$ is implied due to the requirement of
gauge invariance.

As for Fig. 2b, one has to be careful in extracting the $SU(3)$
structure of the resulting contributions to the $Z$-matrix,
because the quartic scalar vertex in this diagram involves both
gauge and superpotential couplings. For holomorphic operators, we
have
\beq{13}
\delta Z^{(b)\cdots  J_lJ_{l+1}\cdots }_{\quad \cdots
I_lI_{l+1}\cdots} =-\frac{\ln{\Lambda}}{8\pi^2}\;\left \{(
\frac{\lambda_{A_l^-}}{2}-\lambda_h)\,
\delta_{I_l}^{J_l}\delta_{I_{l+1}}^{J_{l+1}}
+\lambda_h\delta_{I_l}^{J_{l+1}}\delta_{I_{l+1}}^{J_l} \right \}.
\eeq
For non-holomorphic operators the $Z$-matrix exhibits some
complications:
\beq{14}
\delta Z^{(b)\cdots J_l\bar{J}_{l+1}\cdots}_{\quad \cdots
I_l\bar{I}_{l+1}\cdots} &=& \frac{\ln{\Lambda}}{16\pi^2}
\lambda_{A_l^-}\;
(\delta_{I_l}^{J_l}\delta_{\bar{I}_{l+1}}^{\bar{J}_{l+1}}
+\delta_{I_l,\bar{I}_{l+1}}\delta^{J_l,\bar{J}_{l+1}}), \nonumber \\
\delta Z^{(b)\cdots \bar{J}_lJ_{l+1}\cdots }_{\quad \cdots
I_l\bar{I}_{l+1}\cdots } &=& -\frac{\ln{\Lambda}}{8\pi^2}
\left\{\lambda_h \delta_{I_l}^{J_{l+1}}
\delta_{\bar{I}_{l+1}}^{\bar{J}_l}-(\lambda_h-\frac{\lambda_{A_l^-}}{2})\,
\delta_{I_l,\bar{I}_{l+1}}\delta^{\bar{J}_l,J_{l+1}}\right\},
\nonumber \\ \delta Z^{(b)\cdots \bar{J}_lJ_{l+1}\cdots
}_{\quad\cdots \bar{I}_lI_{l+1}\cdots } &=& \frac{\ln{\Lambda}}
{16\pi^2} \lambda_{\bar{A}_l^+}\;
(\delta_{\bar{I}_{l}}^{\bar{J}_{l}}\delta_{I_{l+1}}^{J_{l+1}}
+\delta_{\bar{I}_l,I_{l+1}}\delta^{\bar{J}_l,J_{l+1}}),
\nonumber \\
\delta  Z^{(b)\cdots J_l\bar{J}_{l+1}\cdots }_{\quad \cdots
\bar{I}_lI_{l+1}\cdots} &=& -\frac{\ln{\Lambda}}{8\pi^2}\left\{
\lambda_h\delta_{\bar{I}_l}^{\bar{J}_{l+1}}
\delta_{I_{l+1}}^{J_l}-(\lambda_h-
\frac{\lambda_{\bar{A}_l^+}}{2})\, \delta_{\bar{I}_l,I_{l+1}}
\delta^{J_l,\bar{J}_{l+1}}\right\}.
\eeq

In next section, we will consider the operator mixing resulting
from above $Z$-factors.

\section{Operator mixing and spin chain}

The anomalous dimension matrix (ADM) at one loop for operators~
(\ref{7}) is determined by the standard arguments through
\beq{15}
\Gamma=\frac{d\delta Z}{d\ln{\Lambda}}.
\eeq
Operator mixing arises when one diagonalizes this matrix to obtain
operators with definite dimension. The ADM acts on a
$18^L$-dimensional vector space ${\cal V}_1\otimes...\otimes {\cal
V}_L$, where each ${\cal V}_l$ is a complex vector space spanned
by $\phi_A^a$ and $\phi_{\bar A}^{\bar a}$, As proposed by Minahan
and Zarembo \cite{MZ02} for the $\CN{4}$ SYM, it is extremely
useful to identify the ADM as the Hamiltonian of a lattice spin
system, where the lattice sites are labeled by the subscript
$l=1,2,\cdots,L$. The main goal of our paper is to look for an
integrable spin chain for the ADM, at least in a sector, in our
model.

However, there is a technical complication here. In the ${\cal
N}=4$ case, the adjoint scalars are labeled only by one index
$i=1,\cdots,6$ for the $SO(6)$ symmetry. But in our model, we need
a pair of indices  $(a,A)$ (or $\bar{a},\bar{A}$) to label the
bi-fundamental scalars, where the extra index $A$ is necessary to
label the arrows in the quiver diagram, indicating which pair of
$U(N)$ gauge groups the scalar is coupled to. This is because our
model is a deformation of an orbifolded model. We note, however,
that in accordance with the quiver diagram, Fig. 1, it is easy to
see that gauge invariance of a composite operator (\ref{7})
strongly constrains its index sequence $I_1,\cdots,I_L$. For
example, if $I_l$ is known, then gauge invariance dictates the
gauge index in $I_{l+1}$, depending on whether its $SU(3)$ index
is of type $a$ or type $\bar{a}$. Thus if $A_1$ is known, then
gauge index sequence of a gauge invariant composite operator is
completely determined by its $SU(3)$ index sequence, corresponding
to an $SU(3)$ spin chain, with spin on each site belongs to either
${\bf 3}$ or $\bf{\bar{3}}$. The Hilbert space at each site is a
6-dimensional (reducible) representation space $V$ of $SU(3)$:
$V={\bf 3}\oplus {\bf \bar{3}}$, and the dimensionality of the
Hilbert space of the spin chain is thus reduced to $6^L$.

In order to write ADM of operators~(\ref{7}) in compact form, and
to facilitate later comparison with the $\CN{4}$ mother SYM
theory, we introduce the projection operators $J^{\pm}$, which
projects $V$ to its invariant sub-spaces ${\bf 3}$ and ${\bf
\bar{3}}$, respectively; in component form, they projects a vector
$v^i\equiv (v^a,\bar{v}^{\bar a})$ to its components $v^a$ and to
$\bar{v}^{\bar a}$. We also define the permutation operator $P$
and the trace operator $K$, which act on the tensor product
$V\otimes V$, respectively, as
\beq{16}
P(u\otimes v)&=&v\otimes u,\nonumber \\
K(u\otimes v)&=&(u\cdot v)\sum_a (\hat{e}_a \otimes \hat{e}_{\bar
a} +\hat{e}_{\bar a} \otimes \hat{e}_a),
\eeq
where $u,\;v$ are vectors in $V$ and $(u \cdot v) = \sum_a (u^a
v^{\bar a} + u^{\bar a} v^a)$; moreover, $\hat{e}_a$ and
$\hat{e}_{\bar b}$ are vectors of an orthonormal basis in $V$.

Then we write the ADM defined in Eq.~(\ref{15}) as follows:
\beq{17}\Gamma_{A_1\cdots A_L}&=&\sum_{l=1}^L\sum_{i,j=\pm}
\Gamma_{l,l+1}^{ij}(A_l)J^i_lJ^j_{l+1}, \nonumber \\
\Gamma_{l,l+1}^{++}(A_l)&=&\Gamma_{l,l+1}^{--}(A_l)=-\gamma_{A_l}
+\frac{\lambda_h}{8\pi^2}(1-P_{l,l+1}), \\
\Gamma_{l,l+1}^{+-}(A_l)&=&-\gamma_{A_l}
+\frac{\lambda_{A_l^+}}{16\pi^2}\sum_{l=1}^L
(2+K_{l,l+1}-K_{l,l+1}P_{l,l+1})
-\frac{\lambda_h}{8\pi^2}\sum_{l=1}^L(1-K_{l,l+1})P_{l,l+1},
\nonumber \\
\Gamma_{l,l+1}^{-+}(A_l)&=&-\gamma_{A_l}
+\frac{\lambda_{A_l^-}}{16\pi^2}\sum_{l=1}^L
(2+K_{l,l+1}-K_{l,l+1}P_{l,l+1})
-\frac{\lambda_h}{8\pi^2}\sum_{l=1}^L(1-K_{l,l+1})P_{l,l+1}.
\nonumber
\eeq
We note that though the coefficients on the right side depend on
the gauge indices $A_l$, but the operators $P_{l,l+1}$ and
$K_{l,l+1}$ act only on the $SU(3)$ indices.

Since at planar one-loop level holomorphic (or anti-holomorphic)
operators, consisting of ${\bf 3}$'s (or of ${\bf{\bar 3}}$'s,
respectively), mix only among themselves. So it makes sense to
restrict the ADM to the holomorphic (or anti-holomorphic) sector,
and then to ask whether the so-restricted ADM can be identified
with the Hamiltonian of an integrable spin chain or not. (From the
quiver diagram, one can easily see that gauge invariance requires
(anti-)holomorphic operators have length $L=3k$ with integer $k$.)

From above equations we see that the ADM $\Gamma_{\rm h}$,
restricted in the holomorphic sector and associated with the
projection operators $J^-_1\cdots J^-_L$, is given by \beq{18}
\Gamma_{{(\rm h)}A_1\cdots A_L}=-\sum_{l=1}^L\gamma_{A_l}
+\frac{\lambda_h}{8\pi^2}\sum_{l=1}^L(1-P_{l,l+1}).
\eeq
A same expression can be obtained for anti-holomorphic operators.

It is worthy to note that in Eq. (\ref{18}), the dependence on
gauge indices $A_l$ can actually be eliminated, since gauge
invariance requires that $L=3k$ and the gauge index sequence goes
around the triangular quiver diagram, Fig. 1, only in one
direction. Therefore we can rewrite the ADM in the holomorphic
sector as
\beq{18a}
\Gamma_{(\rm h)}&=&\Gamma_0+\Gamma_1, \nonumber\\
\Gamma_0&=&-\frac{L}{24\pi^2}(\sum_{A=1}^3\lambda_A
-3\lambda_h), \nonumber \\
\Gamma_1&=&\frac{\lambda_h}{8\pi^2}\sum_{l=1}^L (1-P_{l,l+1}).
\eeq
It is easy to see that $\Gamma_0$ is a constant, while $\Gamma_1$
depends only on $\lambda_h$, the superpotential coupling. By
introducing the spin operators
\beq{spin1}S_{ij}^{ab}=\frac{1}{\sqrt{2}}(\delta^a_i\delta^b_j
-\delta^a_j\delta^b_i),
\eeq
for each lattice site the ADM, $\Gamma_1$, can be rewritten in
terms of manifest spin-spin interactions:
\beq{spin2}\Gamma_1&=&\frac{\lambda_h}{8\pi^2}
\sum_{l=1}^L \{S_l^{ab}S_{l+1}^{ab}-(S_l^{ab}S_{l+1}^{ab})^2\}.
\eeq

Since only the permutation operator $P$ appears in the ADM
(\ref{18a}), it can be identified to be the Hamiltonian of an {\it
integrable} spin chain with $SU(3)$ symmetry, for {\it arbitrary}
't Hooft couplings $\lambda_A$. The detail on the integrability of
$SU(M)$ spin chain will be presented in appendix A. The Hilbert
space of this spin chain is the tensor product $U_1\otimes\cdots
\otimes U_L$ with $U_l=\mathbb{C}^3$ spanned by covariant vectors
of $SU(3)$. Thus, though the full ADM (\ref{17}) in our model does
not in general correspond to an integrable spin chain, the ADM
(\ref{18a}) in the holomorphic sector {\it does}.

It is interesting to note that in our model, an integrable $SU(3)$
spin chain (\ref{18a}) appears even if the theory is away from the
conformal fixed line, independent of the values of $\lambda_A$ and
$\lambda_h$. The integrability of the system enable us to find the
exact one-loop anomalous dimensions of (anti-)holomorphic
operators via applying Bethe ansatz equations. Some examples and
computational details are shown in appendix B.

In Sec. II we have shown that at IR the gauge interaction and
superpotential interaction are decoupled. In the limit
$\lambda_h\to 0$, the ADM~(\ref{18a}) is proportional to identity
operator so that (anti-)chiral operators (\ref{7}) do not mix with
each other under planar one-loop renormalization. However, the
case with $\lambda_A\to 0$ is more interesting. When $\lambda_A\to
0$, the present $\net$ gauge theory actually approaches to an
$\CN{1}$ Wess-Zumino model. Then, according to the above
Eq.~(\ref{18a}), the planar one-loop ADM for composite operators
consisting of purely (anti-) chiral scalars in this Wess-Zumino
model also corresponds to an integrable $SU(3)$ spin chain. (Since
the chiral supermultiplets are taken to be bi-fundamental
representations of three global $U(N)$ groups, here the planar
limit makes sense as the limit in which $N\to \infty$ with $hN^2$
kept fixed.) We note that in this Wess-Zumino model there is no
gauge invariance constraint, so the length $L$ of the holomorphic
composite operators for the integrable ADM (or quantum spin chain)
do {\it not} need to be restricted to a multiple of three!

To conclude, we make the following remark: Though in the above we
have argued that the $SU(3)$ spin chain arising from the ADM in
the holomorphic sector is "blind" to the discrete index $A$, the
operator mixing is {\it not}.  Because each $\phi$-field factor in
the composite operators now carries the index $A$ due to
orbifolding, as we will see in appendix B, the operator mixing
becomes completely different from that for the $SO(6)$ spin chain
in $\CN{4}$ SYM.

\section{Generalization}

\subsection{Lift to the ${\cal N}=4$ SYM}

We have mentioned that on the conformal fixed line, our present
$\CN{1}$ model is a $\mathbb{Z}_3$ orbifolding of ${\cal N}=4$ SYM
\cite{KS98,LNV98}. Then all correlation functions of the
orbifolded theory are known to coincide with those of ${\cal N}=4$
SYM, modulo a rescaling of the gauge coupling constant. This has
been shown either by using string theory \cite{BZV98}, or by using
Feynman diagrams in field theory \cite{BJ98}. Thus we expect that
the spin Hamiltonian~(\ref{18a}) is related to the integrable
$SO(6)$ spin Hamiltonian obtained by Minahan and Zarembo
\cite{MZ02}.

To see this without explicit calculation, we may argue as follows.
First notice that for fixed index $A$, the bi-fundamental $\phi^a$
fields, together with their conjugate $\phi_{\bar{a}}$, form a
6-dimension vector. They are originated from the $\mathbb{Z}_3$
orbifold projection of the 6-dimensional anti-symmetric
representation of $SU(4)$ R-symmetry in ${\cal N}=4$ SYM:
\beq{19}\phi^a=\gamma_g^\dag({(R_g^{\bf
6})^a}_b\phi^b)\gamma_g,\hspace{0.5in}{\rm for\;\; any\;\;}
g\in\mathbb{Z}_3.
\eeq
Here $\gamma_g$ are the regular representation of
$g\in\mathbb{Z}_3$ in $U(3N)$ group, $R_g^{\bf 6}$ the
6-dimensional representation of $g\in\mathbb{Z}_3$ in $SU(4)$
R-symmetry group. In particular, $\phi^a$ corresponds charge-$2/3$
fields in $\bf{6}$ of $SU(4)$ under $R_g$ transformation, while
$\phi^{\bar{a}}$ charge-$(-2/3)$ ones. According to a statement in
ref. \cite{BJ98}, the $\gamma_g$-action on correlation
functions~(\ref{9}) is trivial in the planar limit. Moreover,
gauge invariance of a composite operator (\ref{7}) requires
$|L-2k|/3$ to be integer, where $k$ is the number of
$\phi^{\bar{a}}$. Consequently such composite operators have zero
charge under $R_g$, and the ADM or spin-chain
Hamiltonian~(\ref{17}) at the conformal points should be the same
as the integrable $SO(6)$ spin Hamiltonian in $\CN{4}$ SCYM.

Explicitly on the conformal fixed line, one has
$\lambda_A=\lambda_h=\lambda$, then the ADM~(\ref{17}) is
simplified to
\beq{20}\Gamma&=&
\frac{\lambda}{8\pi^2}\sum_{l=1}^L(1-P_{l,l+1})
(J^+_lJ^+_{l+1}+J^-_lJ^-_{l+1}) \nonumber \\
&+&\frac{\lambda}{16\pi^2}\sum_{l=1}^L
(2+K_{l,l+1}+K_{l,l+1}P_{l,l+1}-2P_{l,l+1})
(J^+_lJ^-_{l+1}+J^-_lJ^+_{l+1}) \nonumber \\
&=&\frac{\lambda}{16\pi^2}
\sum_{l=1}^L(2+\bar{K}_{l,l+1}-2P_{l,l+1}).
\eeq
Here in the last line, $\bar{K}\;(a\otimes b) =(a^\dag\cdot
b)\sum_i\hat{e}^i\otimes\hat{e}^i$. The rank-2 anti-asymmetric
representation of $SU(4)$ is related to the vector representation
of $SO(6)$ via an unitary transformation $U$. Notice that
$(U\otimes U)\bar{K}(U\otimes U)^\dag=K$, the ADM~(\ref{20}) is
indeed the Hamiltonian of the same integrable SO(6) spin chain as
in ref. \cite{MZ02}.

\subsection{Other $\CN{1}$ SYM and Wess-Zumino model}

We have shown that, in our present $\net$ model the ADM of
(anti-)holomorphic composite operators corresponds to a
Hamiltonian of integrable $SU(3)$ spin chain. One may wonder
whether this result can be generalized to other supersymmetric
models with global $SU(M)$ symmetry, since the integrability
condition for a holomorphic spin chain with any $SU(M)$ symmetry
is quite simple, involving only the identity and permutation
operators.

For a general $\CN{1}$ gauge theory with $M$ chiral superfields,
our basic observation is the following: The contributions of
diagrams Fig 2a and 2c are always proportional to the identity
matrix in $SU(M)$ indices. The quartic scalar vertex in Fig 2b
from the gauge interaction (the first term of the Lagrangian~
(\ref{2})) can be effectively treated as an exchange of the
$D$-component of the vector superfield, so its contribution is
also proportional to the identity matrix in $SU(M)$ indices. The
non-trivial $SU(M)$ structure in planar one-loop ADM can only come
from the quartic scalar vertex in Fig. 2b with superpotential
interactions (the third term in eq. (\ref{2})).

Let us first consider a $\net$ $SU(N)$ gauge theory, which has $M$
chiral superfields transforming as ${\bf M}$ of a global $SU(M)$
group and as a certain representation $\bf{\cal R}$ under gauge
group. In order to large $N$ limit makes sense for superpotential
coupling yet, we restrict the representation $\bf{\cal R}$ in
which degrees of freedom of chiral superfields are order to $N^2$
at least. In addition to gauge invariance, the superpotential is
required to be $SU(M)$-invariant and of degree three (for
renormalizability). Therefore the superpotential has to contain an
$SU(M)$-invariant tensor of rank 3. Consequently the only
possibility is $M=3$ and the superpotential contains the
$SU(3)$-invariant tensor $\ep_{abc}$. It is not hard to see that
at the the planar one-loop level, the ADM of (anti-)holomorphic
composite operators contains only the identity and permutation
operator $P$ and, therefore, can be identified with the
Hamiltonian of an integrable $SU(3)$ spin chain.

Now let take the limit in which the 't Hooft coupling tends to
zero, then this model becomes a Wess-Zumino model with three
chiral superfields, each transforming as representation ${\cal R}$
of $SU(N)$ group and together as ${\bf 3}$ of $SU(3)$ flavor
group. With an $SU(3)$-invariant superpotential, the ADM of the
(anti-)holomorphic composite operators again correspond to an
integrable $SU(3)$ spin chain, in the same way as the limit we
discussed at the end of Sec. IV.

\subsection{An $\CN{2}$ SYM}

If the SYM has an extra $U(1)$ R-symmetry, the story will become a
little bit complicated. As an example, let us consider the
orbifolded ${\cal N}=2$ model proposed in \cite{MRV02}. The gauge
group of the model is $SU(N)^{(1)}\times SU(N)^{(2)}\cdots
SU(N)^{(K)}$. The bosonic fields in the vector multiplet are
denoted as $(A_{\mu I},\phi_I)$ with $I=1,\cdots,K$. The matter
fields are hypermultiplets $B_I^a,\;(a=1,2)$, where $B_I$ belongs
to the bi-fundamental representation $(N^{(I-1)},\bar{N}^{(I)})$
of the $(I-1)$-th and $I$-th gauge groups, and they form a doublet
(labeled by the index $a$) under R-symmetry $U(1)_R\times
SU(2)_R$. In order to construct a renormalizable, gauge-invariant
and $SU(2)_R$-invariant superpotential, one has to requires $K=2$.
The superpotential is of the form
\beq{21}{\cal W}=\frac{h}{2}\ep_{ab}{\rm tr}
(B_1^aB_2^b\Phi_1+B_2^aB_1^b\Phi_2),
\eeq
where $a,b=1,2$ are $SU(2)_R$ indices.

We can form three types of gauge-invariant, holomorphic composite
operators ${\cal O}_i,\;(i=1,2,3)$ with $L$ sites, which are
closed under planar one-loop renormalization: ${\cal O}_1$
consisting of the doublet $B_I$s only, ${\cal O}_2$ of $\phi_I$
only and ${\cal O}_3$ with mixed $B_I$ and $\phi_I$. The closure
and holomorphy of ${\cal O}_2$ restricts one-loop planar diagram
corrections to ${\cal O}_2$ to be always diagonal. Its anomalous
dimension is
\beq{22}\Gamma_{2,I}=\frac{L}{8\pi^2}
(\lambda_h-\lambda_I),
\eeq
where $\lambda_I$ denotes the 't Hooft coupling of the $I$th gauge
group. It is interesting to note that at conformal points,
$\lambda_I=\lambda_h$, the operator ${\cal O}_2$ is protected.

The $\ep_{ab}$ in the superpotential~(\ref{21}) forbids the trace
operator $K$ to appear in the ADM of operators ${\cal O}_1$.
Moreover, the quartic scalar vertex derived from the
superpotential~(\ref{21}) contains a term
$(B_1^1B_2^2-B_1^2B_2^1)(\bar{B}_2^2\bar{B}_1^1
-\bar{B}_2^1\bar{B}_1^2)$. The permutation operator $P$ appears,
due to this term, in the ADM of ${\cal O}_1$:
\beq{23}\Gamma_{1,I_1}=\frac{1}{8\pi^2}\sum_{l=1}^{L}
(2\lambda_h-\lambda_{I_l}-\lambda_hP_{l,l+1}).
\eeq
Here it should be noticed that $I_l$ ($l>1$) is uniquely
determined by $I_1$ due to gauge invariance. The ADM in eq.
(\ref{23}) can again be regarded as the Hamiltonian of an
integrable spin chain with $SU(2)$ symmetry.

The case for ${\cal O}_3$ is more complicated, since a $\phi_I$
field and a neighboring $B_J$ may exchange in planar one-loop
diagrams. Effectively we can treat an ${\cal O}_3$ operator as
insertions of $\phi$'s in a ${\cal O}_1$ operator. Noticing an
${\cal O}_3$ consists of $L$ $B_I$s and $k$ $\phi$s is closed with
fixed $L$ and $k$, we can write down its ADM explicitly. For
example, let us consider a $\phi_{I_i}$ inserting at $i$-th site
of ${\cal O}_1$, so that we get operator ${\cal O}_3$ with $L+1$
sites which consists of $L$ $B_I$s and a $\phi$. Its ADM can be
written as \beq{24}\Gamma_{3,(I_1\cdots I_i\cdots I_L)}
=\frac{1}{8\pi^2}\sum_{l=1}^{L+1} (2\lambda_h-\lambda_{I_l})
-\frac{\lambda_h}{8\pi^2}\sum_{l=1,\atop l\neq
i-1,i}^{L+1}P_{l,l+1}-\frac{\lambda_h}{8\pi^2}
(P'_{i-1,i}+P'_{i,i+1}),
\eeq
where the permutation operator $P'$ exchange $\phi_{I_i}$ and
neighboring $B_I$.

We can see the ADM~(\ref{24}) is similar to the ADM~(\ref{23}), if
we suppress the gauge group indices. Therefore, we can formally
define a Hilbert space ${\cal H}=\prod_{l=1}^L\otimes V$ with
$V=\mathbb{C}^2\oplus \mathbb{C}$. That is, to put $B_I^a$ and
$\phi_I$ into a triplet and to consider the set of gauge invariant
operators $\{{\cal O}\} =\{{\cal O}_1\}\oplus\{{\cal O}_2\}
\oplus\{{\cal O}_3\}$. The ADM for this set of operators is
\beq{25}\Gamma_{I_1}=\frac{1}{8\pi^2}\sum_{l=1}^{L}
(2\lambda_h-\lambda_{I_l}-\lambda_hP_{l,l+1}).
\eeq
The above ADM can be regarded as the Hamiltonian of an integrable
spin chain with $SU(3)$ (instead of $SU(2)$) symmetry. It implies
that in the ADM of holomorphic composite scalar operators, its
symmetry (as spin chain) is enhanced from $SU(2)$ $R$-symmetry to
$SU(3)$, if gauge group indices are suppressed, as allowed by
gauge invariance constraint. The difference between the
Hamiltonian~(\ref{25}) and (\ref{18a})is only that they have
different constant terms.

Finally, since this ${\cal N}=2$ model is obtained by deforming an
$\mathbb{Z}_2$ orbifolding of ${\cal N}=4$ SYM, the ADM for gauge
invariant composite scalar operators coincides with that in ${\cal
N}=4$ SYM at conformal points. The same as in the $\net$ case, it
corresponds to the Hamiltonian of an integrable spin chain with
$SO(6)$ symmetry. Consequently for the ADM of SYM theories we
obtain a cascade of integrable structures from orbifolding (or
taking quotient of) ${\cal N}=4$ SYM (see Fig. 3).

\begin{figure}[hptb]
\label{f4} \centering
\includegraphics[width=4in]{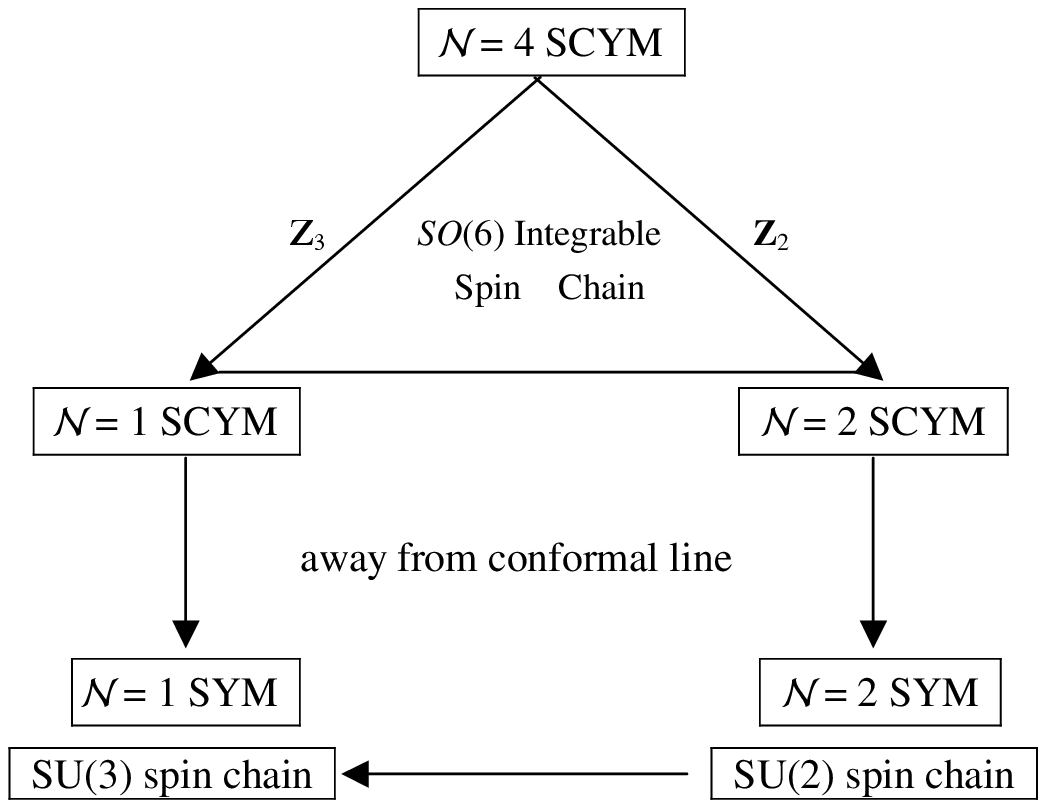}
\begin{minipage}{5in}
\caption{The cascade of integrable structure in SYM.}
\end{minipage}
\end{figure}

\subsection{A conjecture}

In the above examples we have seen that in these orbifolded
daughters of $\CN{1,2}$ SYM, the variation of 't Hooft couplings
affects only the total sum of constant terms in the Hamiltonian of
the quantum spin chain, and the change by an overall factor in the
superpotential couplings only leads to an overall factor for the
spin chain coupling. Neither of them affect the integrability of
the spin chain Hamiltonian. So it is the symmetry structure of the
superpotential that dictates the integrable structure and,
moreover, the integrable structure should exist for all deformed
orbifolded daughters of $\CN{4}$ SYM with unbroken
supersymmetries.

More precisely, we are led to make the following conjecture: In
{\it all} $\CN{1,2}$ orbifolded daughters \cite{KS98,LNV98} of the
$\CN{4}$ SYM, (with a non-abelian global symmetry), with their 't
Hooft and superpotential couplings deformed away from the
conformal points while keeping the global symmetry of the
superpotential, there is always a nontrivial {\it integrable
structure} in operator mixing at planar one-loop level for (anti-)
holomorphic composite operators, at least for those consisting of
purely scalars without derivatives. Moreover, this integrable
structure survives in the resulting Wess-Zumino models when all 't
Hooft couplings are sent to zero.

\section{Summary}

We have presented examples in which an integrable structure
appears in four dimensional $\net,2$ super Yang-Mills theories or
Wess-Zumino models, even with conformal symmetry broken. In these
examples the planar one-loop anomalous dimension matrix for
composite operators consisting of (anti-)holomorphic scalars can
be written as the Hamiltonian of an integrable spin chain with
$SU(3)$ (or $SU(2)$) symmetry. It indicates that the origin of the
integrable structure in four dimensional Yang-Mills theories is of
more profound origin, not restricted to superconformal symmetry.
Though these spin chains can be formally viewed as a subsector of
the integrable $SO(6)$ chain found in $\CN{4}$ SCYM\cite{MZ02}, as
will be shown in appendix B, the operator mixing in our models is
completely different, because each scalar field factor in the
composite operators now carries an extra discrete index $A$, which
is absent in $\CN{4}$ SYM.  Moreover, the appearance of an
integrable structure in various SYM theories enables us to use
methods in one-dimensional integrable models to study properties
of SYM. In appendix A and B below, we will show some examples for
how to find one-loop anomalous dimensions of composite operators
of SYM via solving Bethe ansatz equations.

Though the study of the ADM of composite operators in $\CN{4}$ SYM
was first motivated by the BMN limit, which is conjectured to be
dual to string theory in the pp-wave background, the integrability
of the ADM, as a spin chain Hamiltonian, has nothing to do with
the BMN limit. All our examples confirm this point once more.

As we have seen both in the $\CN{4}$ SYM and our models with
$\CN{1,2}$, the length of the integrable spin chain can be finite.
Moreover we have shown that non-trivial integrable structure of
these spin chain is actually encoded in the superpotential
interactions, rather than gauge interactions. In the cases we
studied, it is the symmetry of the superpotential that determines
the symmetry of the integrable spin chain, and it is the
superpotential coupling appears in the non-trivial part of the
integrable Hamiltonian, even though our deformation of the
superpotential strength has made the theory away from the
conformal points.

Recently there are further developments on the $\CN{4}$ SYM
integrable super spin chain: A study on the two-loop corrections
to the integrable Hamiltonian yielded intriguing evidence that the
higher order corrections do not break the integrability
property\cite{BKS03}. In addition, it was shown that there is a
relation between the infinite-dimensional non-local symmetry of
type IIB superstring in $AdS_5\times S^5$ and a non-abelian and
nonlinear infinite-dimensional Yangian algebra for weakly coupled
SCYM \cite{DNW03,AS03}. It would be interesting to address similar
issues in the framework of $\net,2$ supersymmetric models, such as
those we have presented in this paper. Moreover, from the string
theory point of view, the following questions arise naturally: Is
some of the gauge theories studied here dual to gravity or string
theory? Could the integrable structure in the supersymmetric gauge
theories be originated from or related to string theory?

\begin{acknowledgments}
Both authors thank the Interdisciplinary Center for Theoretical
Study, Chinese Academy of Sciences in Beijing for support, and the
Institute of Theoretical Physics, Chinese Academy of Sciences for
warm hospitality. Ting-liang Zhuang has participated in an early
stage of this work. This work is partly supported by the NSF of
China, Grant No. 10305017, and through USTC ICTS by grants from
the Chinese Academy of Science and a grant from NSFC of China.
\end{acknowledgments}

\appendix

\section{Generalized Heisenberg (anti-)ferromagnet}

We will show that the spin Hamiltonian in Eq.~(\ref{18a}) is just
one of the generalized Heisenberg ferromagnet models discussed in
ref.\cite{KR81}. For readers' convenience, in this appendix we
give a brief review on the generalized Heisenberg
(anti-)ferromagnet --- a quantum system with $M$ components on a
one-dimensional chain with nearest-neighbor (short-range)
interactions. The complete space of state is
\beq{a1}{\mathcal
H}=\bigotimes_{l=1}^L {\cal H}_l,\hspace{0.5in}{\cal
H}_l=\mathbb{C}^M.
\eeq
The Lax operator, associated with the $n$-th site on the chain,
$L_{n,a}(\mu)$, acts on the tensor space ${\cal H}_n\otimes V_a$
with auxiliary space $V_a=\mathbb{C}^M$. It is of the form
\beq{a2}L_{n,a}(\mu)=a(\mu)I_{n,a}+b(\mu)P_{n,a}, \eeq where
$P_{n,a}$ is the permutation operator acting on ${\cal H}_n\otimes
V_a$, $\mu$ the spectral parameter, and $a(\mu)+b(\mu)=1$,
$a(\mu)=\mu/(\mu+i\vep)$ with $\vep=\pm 1$ corresponding to
anti-ferromagnet and ferromagnet respectively.

The transfer matrix $T_{L,a}(\mu)$ defined by
\beq{a3}T_{L,a}(\mu)=L_{L,a}(\mu)\cdots L_{1,a}(\mu)
\eeq
is a monodromy around a circle (assuming the periodical boundary
conditions with ${\cal H}_{L+1}={\cal H}_1$ for the chain). It
satisfies the following relations
\beq{a4}R_{ab}(\mu-\nu)T_{L,a}(\mu)T_{L,b}(\nu)=
T_{L,b}(\nu)T_{L,a}(\mu)R_{ab}(\mu-\nu),
\eeq
where $R_{ab}(\mu)=b(\mu)+a(\mu)P_{a,b}$. Taking the trace on the
auxiliary spaces $V_a$ and $V_b$, we get the commutative relation
\beq{a5}[t(\mu),t(\nu)]=0, \hspace{1in}t(\mu)={\rm
tr}_a\;T_{L,a}(\mu).
\eeq
This allows us to treat $t(\mu)$ as a generating function of
commuting conserved quantities:
\beq{a6}M^{(l)}=i\left(\frac{d}{d\mu}\right)^l\;
\ln{[t(\mu)t(0)^{-1}]}|_{\mu=0}.
\eeq
By definition, the momentum operator $P$ on a lattice,
$P=-i\ln{t(0)}$, is related to the discrete shift operator (by one
site)
\beq{a7}t(0){\cal H}_1\otimes{\cal H}_2\otimes\cdots\otimes
{\cal H}_L={\cal H}_2\otimes\cdots\otimes {\cal H}_L\otimes {\cal
H}_1.
\eeq
While the Hamiltonian of the system is
\beq{a8}H=M^{(1)}=\vep\sum_{l=1}^L(P_{l,l+1}-1).
\eeq
Then we can see that for very large $L$ the spin
Hamiltonian~(\ref{18a}) for our $\CN{1}$ model is nothing but
Eq.~(\ref{a8}) with $\vep=-1$ (for a ferromagnetic system).

For ${\cal H}_n=\mathbb{C}^3$, the eigenvalues $\Lambda(\mu)$ of
the operator $t(\mu)$ is \cite{KR81}
\beq{a9}\Lambda(\mu)=a(\mu)^L\left\{\prod_{j=1}^m\frac{1}
{a(\mu-\mu_j^{(1)})}+\prod_{i=1}^n\frac{1}{a(\mu-\mu_i)}
\prod_{j=1}^m\frac{1}{a(\mu-\mu_j^{(1)})}\right\}
+\prod_{i=1}^n\frac{1}{a(\mu_i-\mu)}.
\eeq
Here the rapidity variables $\mu_i$ and $\mu_j^{(1)}$ satisfy the
algebraic Bethe Ansatz equations (ABAE):
\beq{a10}
\prod_{k=1}^n a(\mu_j^{(1)}-\mu_k)&=& \prod_{k=1\atop k\neq j}^m
\frac{a(\mu_j^{(1)}-\mu_k^{(1)})}{a(\mu_k^{(1)}-\mu_j^{(1)})},
\nonumber \\
a(\mu_j)^L\prod_{k=1}^m\frac{1}{a(\mu_k^{(1)}-\mu_j)}&=&
\prod_{k=1\atop k\neq j}^n \frac{a(\mu_j-\mu_k)}{a(\mu_k-\mu_j)}.
\eeq
The eigenvalues of the momentum operator $P$ and Hamiltonian $H$
are, respectively,
\beq{a11}p(\{\mu_j\})&=&\sum_{j=1}^np(\mu_j)=\frac{1}{i}\sum_{j=1}^n
\ln{\frac{\mu_j+i\vep}{\mu_j}}, \nonumber \\
E(\{\mu_j\})&=&\sum_{j=1}^n\ep(\mu_j)=-\sum_{j=1}^n
\frac{\vep}{\mu_j(\mu_j+i\vep)}.
\eeq
Introducing new rapidity variables $\mu_{1,j}$ and $\mu_{2,j}$ via
\beq{a12}\mu_j&=&\frac{1}{2}\mu_{1,j}-\frac{i}{2}\vep, \nonumber\\
\mu_j^{(1)}&=&\frac{1}{2}\mu_{2,j}-i\vep,
\eeq
we get the usual expressions for the ABAE~(\ref{a10})
\beq{a13}
\left(\frac{\mu_{1,j}-i\vep}{\mu_{1,j}+i\vep}\right)^L&=&
\prod_{k=1\atop k\neq j}^{n_1}\frac{\mu_{1,j}-\mu_{1,k}-2i\vep}
{\mu_{1,j}-\mu_{1,k}+2i\vep}\prod_{l=1}^{n_2}
\frac{\mu_{1,j}-\mu_{2,l}+i\vep} {\mu_{1,j}-\mu_{2,l}-i\vep},
\nonumber \\
1&=&\prod_{k=1\atop k\neq
j}^{n_2}\frac{\mu_{2,j}-\mu_{2,k}-2i\vep}
{\mu_{2,j}-\mu_{2,k}+2i\vep}\prod_{l=1}^{n_1}
\frac{\mu_{2,j}-\mu_{1,l}+i\vep}{\mu_{2,j}-\mu_{1,l}-i\vep},
\eeq
and for eigenvalues of total momentum and energy, we have
\beq{a14}p&=&\sum_{j=1}^{n_1}p(\mu_{1,j}),\hspace{0.5in}
p(\mu)=\frac{1}{i}\ln{\frac{\mu-i\vep}{\mu+i\vep}}, \nonumber\\
E&=&\sum_{j=1}^{n_1}\ep(\mu_{1,j}),\hspace{0.5in}
\ep(\mu)=-\frac{4\vep}{\mu^2+1}.
\eeq

\section{Anomalous dimensions from Bethe Ansatz}

In this appendix we present several concrete examples to show how
to obtain anomalous dimensions by solving Bethe ansatz equations.
The discussion will be similar to that in refs.
\cite{MZ02,Faddeev96}, but there are new issues to address,
associated with the discrete index $A$ arising from orbifolding,
which is absent in the $\CN{4}$ case. In section 4 we have shown
that the ADM for holomorphic operators in our $\CN{1}$ model gives
rise to an $SU(3)$ spin chain, which can be viewed as a closed
sector of the $SO(6)$ spin chain in the $\CN{4}$ SYM. However, as
we will see below, the operator mixing in our model is completely
different, because each $\phi$-field factor now carries an extra
discrete index $A$ that labels the gauge group factors.

The $SU(3)$ symmetry of our spin chain is related to Lie algebra
$A_2$, which has two simple roots,
\beq{b1}\vec{\alpha}_1=(\sqrt{\frac{3}{2}},-\frac{1}{\sqrt{2}}),
\hspace{0.5in} \vec{\alpha}_2=(0,\sqrt{2}),
\eeq
and the highest weight vectors that generate the fundamental and
anti-fundamental representation are, respectively,
\beq{b2}\vec{w}_1=(\sqrt{\frac{2}{3}},0),\hspace{0.5in}
\vec{w}_2=(\frac{1}{\sqrt{6}},\frac{1}{\sqrt{2}}).
\eeq
We always take $\vep=-1$ in the rest of this appendix. Then the
ABAE~(\ref{a13}) follows from the general ABAE \cite{OW86}:
\beq{b3}
\left(\frac{\mu_{q,j}+i\vec{\alpha}_q\cdot\vec{w}}
{\mu_{q,j}-i\vec{\alpha}_q\cdot\vec{w}}\right)^L&=&
\prod_{k=1\atop k\neq j}^{n_q}\frac{\mu_{q,j}-\mu_{q,k}
+i\vec{\alpha}_q\cdot\vec{\alpha}_q}
{\mu_{q,j}-\mu_{q,k}-i\vec{\alpha}_q\cdot\vec{\alpha}_q}
\prod_{q'\neq q}\prod_{l=1}^{n_{q'}}
\frac{\mu_{q,j}-\mu_{{q'},l}+i\vec{\alpha}_q\cdot\vec{\alpha}_{q'}}
{\mu_{q,j}-\mu_{{q'},l}-i\vec{\alpha}_q\cdot\vec{\alpha}_{q'}}.
\eeq

\subsection{Physics}

Each factor in the composite operator corresponds a site in the
spin chain, which is occupied by a scalar field, being one
component of the $SU(3)$ triplet $\phi^a$. For the sake of
convenience we denote them by $(Z,\;Y,\;W)$ in this appendix.

\begin{itemize}
\item[a]] The ground state $\Omega$

Because the eigen-energy of the ferromagnet system (\ref{a8}),
namely $\Gamma_1$ in eq. (\ref{18a}), is non-negative, the ground
state $\Omega$ must have zero eigen-energy and total momentum.
Consequently, $\Omega$ must consist of only one component of
$(Z,\;Y,\;W)$. For convenience we assume it consists of $Z$ only
$^2$\footnotetext[2]{In $\CN{4}$ SYM this state correpsonds to the
ground state of the BMN operator.}. Since there is no impurities
of $\mu_1$ and $\mu_2$, it is not hard to see that
$P_{_\Omega}=E_{_\Omega}=0$.

In our $\CN{1}$ SYM, the anomalous dimension of $\Omega$ is given
by the "zero-point energy" of the system ($\Gamma_0$ in
Eq.~(\ref{18a})):
\beq{b4}\gamma_{_\Omega}=\td{E}_0=
-\frac{L}{24\pi^2}(\lambda_1+\lambda_2+\lambda_3-3\lambda_h).
\eeq
Along the conformal line, $\lambda_h=\lambda_A$, one has
$\gamma_{_\Omega}=0$, so that the dimension of $\Omega$ is indeed
protected by superconformal invariance.

\item[b]] States with impurities

In the present case there are two types of impurities in the spin
chain, labeled by two rapidities: $\mu_{1,j}$ and $\mu_{2,j}$,
which are associated with the two simple roots $\vec{\alpha}_1$
and $\vec{\alpha}_2$ respectively.  The states with impurities
correspond to excitations of $W$ and $Y$ (i.e. the replacements of
some $Z$ by $W$ and/or $Y$) in the ground state $\Omega$. Since
now both $\vec{w}_1-\vec{\alpha}_1$ and $\vec{w}_1-\vec{\alpha}_2$
are also weights, in addition to purely $\mu_1$ or $\mu_2$
impurities, $\mu_1-\mu_2$ bounded impurities are also allowed.
However, the weight $\vec{w}_1-\vec{\alpha}_2$ is not equivalent
to $\vec{w}_1$, so it does not lie in the fundamental
representation. If we restrict ourselves to holomorphic operators,
a single $\mu_2$-impurity without being bounded to a
$\mu_1$-impurity on the same site is not allowed. The physical
interpretation is the following: A single $\mu_1$-impurity
($\vec{w}_1-\vec{\alpha}_1$) creates a $W$ replacement in the
state $\Omega$, while a $\mu_1-\mu_2$ bounded impurity
($\vec{w}_1-\vec{\alpha}_1-\vec{\alpha}_2$) creates a $Y$
replacement. But an individual $\mu_2$-impurity
($\vec{w}_1-\vec{\alpha}_2$) would create a $\bar{Y}$ replacement
in $\Omega$, breaking the holomorphic nature of the composite
operator.
\item[c]] The trace condition

An important observation is that the composite operators that we
are considering are gauge invariant after taking the trace in the
product gauge group $U(N)\times U(N)\times U(N)$, while the scalar
fields belong to bi-fundamental representations of two gauge
groups, instead of the adjoint of one group. So gauge invariance
and holomorphy of the composite operators requires that we are
dealing with a chain with length $L=3k$ with $k$ integer, and
\beq{b5}t(0)^3\Psi(\{\mu\})=\Psi(\{\mu\}).
\eeq
In other words, after shifting the chain by one site three times
in the same direction, we should obtain the same composite
operator. Therefore we have the cubic trace condition:
\beq{b6}
\left(\prod_{j=1}^{n_1}
\frac{\mu_{1,j}+i}{\mu_{1,j}-i}\right)^3=1.
\eeq
It is easy to verify that this trace condition is consistent with
the ABAE~(\ref{a13}). It follows that the total momentum can be
only $2n\pi/3$ with $n$ integer.

We note that the condition (\ref{b5}) or (\ref{b6}) corresponds
to, but is very different from, the trace condition of the $SO(6)$
spin chain in $\CN{4}$ SYM \cite{MZ02}. In the latter case, one
has power 1 instead of power 3 in Eq. (\ref{b5}) and Eq.
(\ref{b6}). The reason for power 3 in our model is obviously
related to the orbifolding by $Z_3$, which leads to an extra
three-valued index $A$ for the holomorphic scalars. Below we will
see that it is the appearance of this index, though the spin
Hamiltonian is "blind" to it, that makes the operator mixing very
different from that in $\CN{4}$ SYM.
\end{itemize}

\subsection{One impurity}

The first non-trivial and interesting case is a single
$\mu_1$-impurity in $\Omega$. This is the simplest example which
shows how and why the operator mixing in our model is very
different from that in the $\CN{4}$ SYM, though the spin
Hamiltonian may be viewed as a closed subsector in the $SO(6)$
spin chain in the latter. In $\CN{4}$ SYM, a single impurity with
non-zero momentum is not allowed, because of the trace condition.
As shown below, however, the present case does allow a single
$\mu_1$-impurity.

The trace condition and the ABAE now reduce to
\beq{b7}\left(\frac{\mu_1+i}{\mu_1-i}\right)^3=1.
\eeq
Here we have used the fact $L=3k$ with $k$ integer. The above
equation yields
\beq{b8}p&=&p(\mu_1)=\frac{2n\pi}{3},\nonumber \\
E&=&\ep(\mu_1)=4\sin^2{\frac{n\pi}{3}}.
\eeq
Then anomalous dimension is
\beq{b9}\gamma_n=\frac{\lambda_h}{2\pi^2}\sin^2{\frac{n\pi}{3}}
+\gamma_{_\Omega}.
\eeq

In the language of $\CN{1}$ SYM, we are now considering
excitations with one $W$ replacement in the ground state $\Omega$.
There are {\it three} distinct possibilities:
\beq{b10}{\cal O}_1&=&{\rm tr}\{W_1Z_2Z_3(Z_1Z_2Z_3)^{k-1}\},
\nonumber \\
{\cal O}_2&=&{\rm tr}\{Z_1W_2Z_3(Z_1Z_2Z_3)^{k-1}\}, \\
{\cal O}_3&=&{\rm tr}\{Z_1Z_2W_3(Z_1Z_2Z_3)^{k-1}\}, \nonumber
\eeq
where the subscripts $1,\;2,\;3$ are values of the index $A$,
labeling three types of bi-fundamentals. In the above operator
basis, the ADM~(\ref{18a}) reads
\beq{b11}\Gamma=\frac{\lambda_h}{8\pi^2}M+\Gamma_0,\hspace{0.5in}
M=\left(\begin{array}{ccc} 2&-1&-1 \\ -1&2&-1 \\ -1&-1&2
\end{array}\right).
\eeq
The matrix $M$ has eigenvalues $\{3,\;3,\;0\}$. Therefore we
obtain the eigenvalues of $\Gamma$ as follows:
\beq{b12}\gamma_1=\gamma_2=\frac{3\lambda_h}{8\pi^2}
+\gamma_{_\Omega},\hspace{0.5in}\gamma_3=\gamma_{_\Omega}.
\eeq
The results are precisely the same as one obtains from the ABAE.
The corresponding eigenvectors are $\{a{\cal O}_1,b{\cal O}_2,
-(a+b){\cal O}_3\}$ for $\gamma_1=\gamma_2$ and $c\{{\cal
O}_1,{\cal O}_2,{\cal O}_3\}$ for $\gamma_3$ with arbitrary
constants $a,\;b,\;c$.

\subsection{Two impurities}

Two impurities can be either two $\mu_1$ or one $\mu_1$ and one
$\mu_2$. Let us consider two $\mu_1$-impurities first, i.e.
excitations with two $W$ replacements in the ground state. The
trace condition and the ABAE now reduce to
\beq{b13}
&&\frac{\mu_{1,1}+i}{\mu_{1,1}-i}\,\cdot\,
\frac{\mu_{1,2}+i}{\mu_{1,2}-i}
=e^{2in\pi/3}, \nonumber \\
&&\left(\frac{\mu_{1,1}+i}{\mu_{1,1}-i}\right)^L=
\frac{\mu_{1,1}-\mu_{1,2}+i}{\mu_{1,1}-\mu_{1,2}-i}.
\eeq

For $n=0$, we must impose $\mu_{1,1}=-\mu_{1,2}$, then only real
solutions are allowed. We get the momenta
$p(\mu_{1,1})=2m\pi/(L-1)$ and the anomalous dimensions (with $m$
an integer):
\beq{b14}\gamma_m^{(n=0)}=\frac{\lambda_h}{\pi^2}\sin^2
{\frac{m\pi}{L-1}}+\gamma_{_\Omega}.
\eeq

For $n=1,2$, however, both of real and complex solutions are
allowed. If $\mu_{1,1}$ is real, from Eq.~(\ref{b13}) we obtain
\beq{b15}
(L-1)\vartheta(\mu_{1,1})=m\pi
+\vartheta(\mu_{1,1}\mp\frac{2}{\sqrt{3}}),\hspace{0.5in}
\mu_{1,2}=\frac{\sqrt{3}\pm\mu_{1,1}}{\sqrt{3}\mu_{1,1}\mp 1},
\eeq
where $\vartheta(x)\equiv \arctan(x)$, the integer $m$
parameterizes different branches of the logarithmic function.
Eq.~(\ref{b15}) in general can not be solved analytically. For
very large $L$, however, we have approximately
$\theta=\vartheta(\mu_{1,1})\simeq m\pi/L$. Consequently the
anomalous dimensions are given by
\beq{b16}\gamma_m^{(n=1,2)}&=&\frac{\lambda_h}{2\pi^2}
(1+\frac{1}{4}\cos{2\theta}\mp\frac{\sqrt{3}}{4}\sin{2\theta})
+\gamma_{_\Omega}\nonumber \\ &=&\frac{\lambda_h}{2\pi^2}
(1+\frac{1}{4}\cos{\frac{2m\pi}{L}}\mp\frac{\sqrt{3}}{4}
\sin{\frac{2m\pi}{L}})+\gamma_{_\Omega}+O(\frac{1}{L^2}).
\eeq

Now let us consider complex solutions. Notice that for
$L\to\infty$, the LHS of the second equation in Eq.~(\ref{b13})
grows (or decreases) exponentially if ${\rm Im}\mu_{1,1}\neq 0$.
Hence the RHS of this equation together with the first equation in
Eq.~(\ref{b13}) lead to the solutions
\beq{b17}\left\{\begin{array}{c}
\mu_{1,1}=\pm\frac{2}{\sqrt{3}}+i, \\
\mu_{1,2}=\pm\frac{2}{\sqrt{3}}-i,
\end{array}\right.,\hspace{0.5in}
{\rm or}\quad \left\{\begin{array}{c}
\mu_{1,1}=\pm\frac{2}{\sqrt{3}}-i, \\
\mu_{1,2}=\pm\frac{2}{\sqrt{3}}+i,
\end{array}\right.,
\eeq
and the anomalous dimension
\beq{b18}\td{\gamma}=\frac{3\lambda_h}{16\pi^2}+\gamma_{_\Omega}.
\eeq
Notice that $\td{\gamma}<{\rm min}(\gamma_m^{(n=1,2)})$. It
indicates that complex solutions correspond to bound states.

Next we consider the bounded impurity of one $\mu_1$ and one
$\mu_2$. The ABAE together with trace condition now reduce to
Eq.~(\ref{b7}) plus
\beq{b19}\frac{\mu_2-\mu_1-i}{\mu_2-\mu_1+i}=1.
\eeq
The solution from those equations are $\mu_1=\cot{\frac{n\pi}{3}}$
and $\mu_2=\infty$. It yields the following anomalous dimension
\beq{b20}\gamma_n=\frac{\lambda_h}{8\pi^2}\ep(\mu_1)
+\gamma_{_\Omega} =\frac{\lambda_h}{2\pi^2}\sin^2{\frac{n\pi}{3}}
+\gamma_{_\Omega}.
\eeq
The result is the same as in Eq.~(\ref{b9}). It just reflects the
fact that replacing a $Z$ in the ground state either by $W$ or by
$Y$ leads to the same anomalous dimension.

\subsection{The highest excited state}

For a finite chain the excited state with the highest energy
contains as many as possible impurities. Taking the logarithm of
the ABAE~(\ref{a13}) we have
\beq{b21}\vartheta(\mu_{1,j})&=&\frac{j\pi}{L}
+\frac{1}{L}\sum_{k\neq j}^{n_1}\vartheta(\frac{\mu_{1,j}
-\mu_{1,k}}{2}) -\frac{1}{L}\sum_{k=1}^{n_2}
\vartheta(\mu_{1,j}-\mu_{2,k}), \nonumber \\ 0&=&\frac{j\pi}{L}
+\frac{1}{L}\sum_{k\neq j}^{n_2}
\vartheta(\frac{\mu_{2,j}-\mu_{2,k}}{2})
-\frac{1}{L}\sum_{k=1}^{n_1}\vartheta(\mu_{2,j}-\mu_{1,k}).
\eeq
where we have used that fact that the discreteness of the Bethe
roots requires them to be pushed to different branches of the
logarithm function. In general the ABAE~(\ref{b21}) can not be
solved analytically with more than two impurities. In the
thermodynamical limit $L\to\infty$, however, the Bethe ansatz
equations are simplified significantly \cite{YY69}. In this limit
$j/L$ is replaced by a continuous variable $x$, and the
ABAE~(\ref{b21}) is replaced by a set of integral equations:
\beq{b22}\vartheta(\mu_1(x))&=&\pi x+\int dy\;
\vartheta(\frac{\mu_1(x)-\mu_1(y)}{2})-\int dy\;
\vartheta(\mu_1(x)-\mu_2(y)), \nonumber \\ 0&=&\pi x+\int dy\;
\vartheta(\frac{\mu_2(x) -\mu_2(y)}{2})-\int dy \;
\vartheta(\mu_2(x)-\mu_1(y)).
\eeq
Taking derivatives with respect to $\mu_1$ and $\mu_2$, we have
\beq{b23}\frac{1}{\mu_1^2+1}&=&\pi\rho_1(\mu_1)
+\int_{-\infty}^\infty
d\mu'\;\frac{2\rho_1(\mu')}{(\mu_1-\mu')^2+4}
-\int_{-\infty}^\infty
d\mu'\;\frac{\rho_2(\mu')}{(\mu_1-\mu')^2+1}, \nonumber \\
0&=&\pi\rho_2(\mu_2) +\int_{-\infty}^\infty
d\mu'\;\frac{2\rho_2(\mu')}{(\mu_2-\mu')^2+4}
-\int_{-\infty}^\infty
d\mu'\;\frac{\rho_1(\mu')}{(\mu_2-\mu')^2+1}, \eeq where the
densities $\rho_1$ and $\rho_2$ are defined by
\beq{b24}\rho_1(\mu_1)=\frac{dx}{d\mu_1(x)},\hspace{1in}
\rho_2(\mu_2)=\frac{dx}{d\mu_2(x)}.
\eeq
The equations~(\ref{b23}) can be solved by means of Fourier
transformation. The results are as follows:
\beq{b25}
\rho_1(x)&=&\int\frac{dk}{2\pi}e^{ikx}\frac{2\cosh{k}}
{4\cosh^2{k}-1}, \nonumber \\
\rho_2(x)&=&\int\frac{dk}{2\pi}e^{ikx}\frac{1}{4\cosh^2{k}-1}.
\eeq
This yields
\beq{b26}\int_{-\infty}^\infty dx\;\rho_1(x)=\frac{2}{3},
\hspace{1in} \int_{-\infty}^\infty dx\;\rho_2(x)=\frac{1}{3}.
\eeq
Consequently we have $2L/3$ $\mu_1$-impurities and $L/3$
$\mu_2$-impurities for the highest excited state. They fill all
sites on the ferromagnet chain. In our $\CN{1}$ SYM, it implies
that there are equal number of $W$, $Y$ and $Z$ scalar fields in
the composite operator. Recalling $L=3k$ with $k$ integer and
denoting the $SU(3)$ triplet as $\phi^a,\;(a=1,2,3)$ again, the
operator has the following form:
\beq{b27}{\cal O}\;\sim\;{\rm tr}(\ep_{abc}\phi^a\phi^b\phi^c)^k.
\eeq
It is an $SU(3)$ singlet with zero total momentum.

The anomalous dimension of this operator is
\beq{b28}\gamma=\frac{\lambda_h}{2\pi^2}
\int_{-\infty}^\infty dx\;
\frac{\rho_1(x)}{x^2+1}+\gamma_{_\Omega}
=\frac{\lambda_h}{24\pi^2}L
(\frac{\pi}{\sqrt{3}}+3\ln{3})+\gamma_{_\Omega}.
\eeq
It grows linearly with $L$, the length of the chain.

Similar to the procedure in \cite{MZ02,Faddeev96}, we can also
calculate the anomalous dimensions of operators with a few $W$ and
$Y$ replacements. But let us stop here.

\end{document}